\documentclass[9pt,twocolumn,twoside]{osajnl}

\journal{ol} 

\setboolean{shortarticle}{true}

\title{Legacy \LaTeX\ template for preparing an article for submission to OSA journals \emph{Applied Optics}, \emph{Advances in Optics and Photonics}, JOSA A, JOSA B, \emph{Optics Letters}, \emph{Optica}, and \emph{Photonics Research}}

\usepackage[]{graphicx}
\usepackage{color}
\usepackage{bm}
\usepackage[english]{babel}
\usepackage{amsmath}
\usepackage{amssymb}
\usepackage{units}
\usepackage{upgreek}
\usepackage[colorlinks=true, allcolors=blue]{hyperref}
\usepackage{ctable}
\usepackage{footnote}
\usepackage{bbold}
\usepackage{siunitx}  
\usepackage{textcomp}
\usepackage{verbatim} 

\title{Coherent beam combining with micro-lens arrays} 

\author[1,*]{Maike Prossotowicz}
\author[2]{Andreas Heimes}
\author[2]{Daniel Flamm}
\author[1]{Florian Jansen}
\author[1]{Hans-J\"{u}rgen Otto}
\author[1]{Aleksander Budnicki}
\author[1]{Alexander Killi}
\author[3]{Uwe Morgner}

\affil[1]{TRUMPF Laser GmbH, Aichhalder Str.\,39, 78713 Schramberg, Germany}
\affil[2]{TRUMPF Laser- und Systemtechnik GmbH, Johann-Maus-Str.\,2, 71254 Ditzingen, Germany}
\affil[3]{Institut für Quantenoptik, Leibniz Universität Hannover, Welfengarten 1, D-30167 Hannover, Germany}

\affil[*]{Corresponding author: maike.prossotowicz@trumpf.com}

\doi{\url{https://doi.org/10.1364/OL.414388}\\ \\ \copyright \,\, \textbf{2020 Optical Society of America. One print or electronic copy may be made for personal use only. Systematic reproduction and distribution, duplication of any material in this paper for a fee or for commercial purposes, or modifications of the content of this paper are prohibited.\\}} 

\begin{abstract}
A novel concept for coherent beam combining is presented based on a simple setup with microlens arrays. These standard components are used in a proof-of-principle experiment for both coherent beam splitting and combination of $5 \times 5$ beams. Here a combination efficiency above $\unit[90]{\%}$ is achieved. We call this novel concept mixed aperture. \end{abstract}

\setboolean{displaycopyright}{true}

\begin{document}

\maketitle
\begin{figure} [t]
  \centering
     \includegraphics[width=0.47 \textwidth]{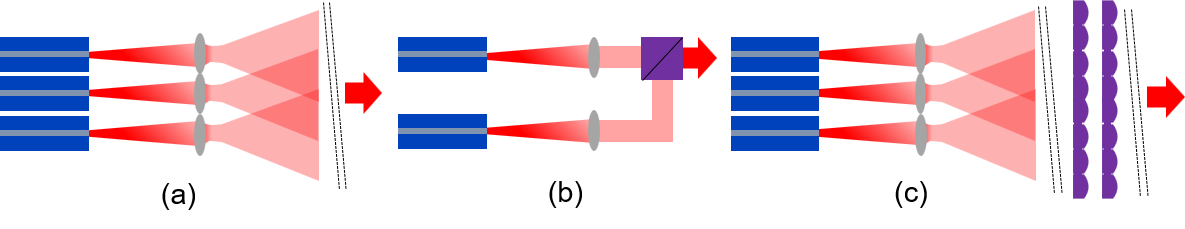}
  \caption{Geometries for CBC. (a) Tiled aperture. (b) Filled aperture. (c) Described geometry in this paper: Mixed aperture.}
  \vspace{.24cm}
  \label{fig:combination_geometry}
\end{figure}

\label{sec:intro}  
High intensity lasers have found many applications \cite{poprawe2004laser} in science and industry. Intrinsic limits, like nonlinear refraction or Raman scattering hinder scaling parameters. Coherent beam combining (CBC) is one of the most promising concepts for further power and energy scaling of laser sources \cite{augst2004coherent,klenke2013530}. With CBC average powers for ultrafast lasers of $\unit[10.4]{kW}$ \cite{muller202010} and pulse energies up to $\unit[12]{mJ}$ \cite{kienel201612} have been demonstrated. This offers a new field of applications for science and industry \cite{ion2005laser},
especially for the processing of materials on large surfaces \cite{tillkorn2018anamorphic} or within large volumes \cite{jenne2020facilitated}.\\
A CBC setup can be considered as an interferometer with several amplifiers. Therefore, laser light of a common seed source is subsequently split and sent into the interferometer arms. In each arm one amplifier is located, so that the split laser light is amplified and combined into a single output beam.\\
For ultrafast lasers, disturbances e.g. atmospheric turbulences or thermal effects cause phase differences between these pulses that require an active compensation. For more details see Ref.\,\cite{hanna2016coherent}.\\ 
With the knowledge of the phase information per channel an error signal is created to match the path length differences between the channels. Therefore, a phase modulator (mirror on a piezo electric actuator, a piezo-electric fiber stretcher or a spatial light modulator (SLM)) must be placed in at least $N-1$ channels. For such a system the combination efficiency depends on the match of the phase differences \cite{hanna2016coherent} and on the combination element itself.\\
In a CBC system the combination geometry is of particular importance, since it has to perform the beam combination while preserving the beam quality and keeping losses as low as possible. Additionally, the optical setup has to resist high average and high peak powers.\\
There are two different types of geometry for the combination setup known, labeled tiled and filled aperture approach (Fig.\,\ref{fig:combination_geometry} (a) and (b)). For the tiled aperture the beams are arranged in the near field side by side \cite{hanna2016coherent} and the combination is achieved in the far field. Therefore no combining element is used \cite{klenke2016performance}. With this geometry theoretical combination efficiencies of $\unit[76]{\%}$ \cite{ramirez2015coherent} are possible. The latest experiments present a combination efficiencies around $\unit[50]{\%}$ \cite{le2017highly} and up to 61 channels \cite{fsaifes2020coherent}.\\ 
For the filled aperture approach the beams are superimposed in near and far field with combining elements \cite{hanna2016coherent,klenke2016performance}, such as partially reflective surfaces \cite{daniault2011coherent}, polarization dependent beam splitters \cite{seise2010coherent} or segmented beam mirrors \cite{klenke2015large}. Hereby combination efficiencies above $\unit[90]{\%}$ \cite{klenke2016performance} are shown.
\begin{figure*} [t]
  \centering
     \includegraphics[width=0.9 \textwidth]{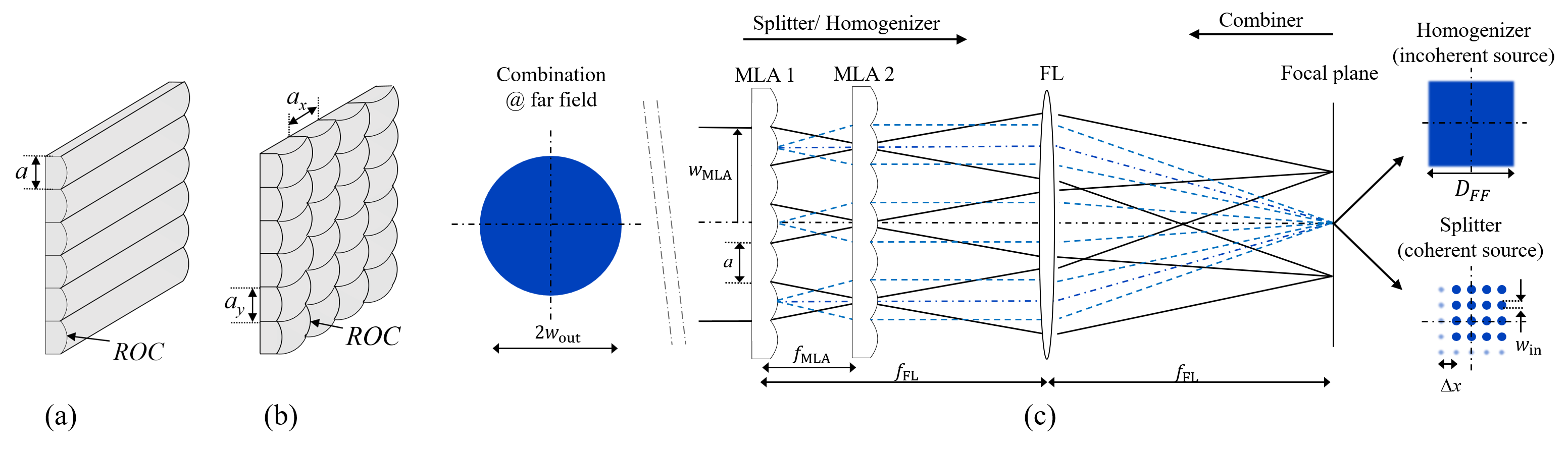}
  \caption{Types of MLAs: (a) cylindrical (b) square MLAs. (c) Setup with MLAs as beam homogenizer, splitter and combiner element.}
  \label{fig:Figure3}
\end{figure*}
\\In this paper a new combining geometry is introduced with one central combining element. Therefore, a pair of micro-lens arrays (MLAs), that is usually used for beam homogenization \cite{tillkorn2018anamorphic,jin2016freeform}, is utilized. The presented novel concept combines the advantages of both well-known combination geometries the tilled and filled aperture approach and works for a low filling factor\,$FF$. From literature it is known, that high quality beam homogenization with MLAs requires a certain reduced level of spatial coherence \cite{zimmermann2007microlens,harder2004homogenization}. A light source exhibiting complete spatial coherence will yield discrete diffraction orders after propagating along the MLA setup. Since the optical power within these diffraction orders can be distributed homogeneously, this approach is used as an efficient beam splitting concept, known already since the early 1990s \cite{streibl1991array}. It is the basic idea of this work to reverse the beam splitting and, thus, to achieve the beam combination. 
\\
MLAs are standard components and used in diverse areas, such as beam homogenization \cite{harder2004homogenization,tillkorn2018anamorphic,zimmermann2007microlens} or for fiber collimation \cite{kikuchi2003fiber}.\\
The principle of the so called imaging homgenizer constitutes the basis for understanding beam splitting and combining. The imaging homogenizer consists of two MLAs, mainly square and cylindrical MLAs (shown in Fig.\,\ref{fig:Figure3} (a) and (b)). The characteristics of these MLAs are the pitch $a$ and radius of curvature $ROC$ which define the effective focal length of a MLA\,$f\textsubscript{MLA}$. The beam combining and splitting for a CBC system with MLAs is based on the theory of an imaging homogenizer \cite{harder2004homogenization} and is shown in Fig.\,\ref{fig:Figure3} (c), where the schematic beam path is presented. In this figure the incident beam illuminates the first MLA and is split into an array of beamlets. The distance between the MLAs is close to $f\textsubscript{MLA}$ and chosen following the Köhler condition \cite{zimmermann2007microlens}. The second MLA in combination with the Fourier-lens\,FL acts as an array of objective lenses that superimposes the images of the generated beamlets of the first MLA in the focal plane of the FL. In geometrical optics a homogeneous intensity profile results.\\ 
With cylindrical MLAs a line profile results and for square MLAs a square profile results in the far field of the MLAs. The width of this square\,$D\textsubscript{FF}$ can be calculated with $D\textsubscript{FF}=a \cdot f\textsubscript{FL}/f\textsubscript{MLA}$ \cite{harder2004homogenization}.\\
In wave optics, coherence leads to interference and grating diffraction \cite{zimmermann2007microlens,harder2004homogenization} and a diffraction pattern results instead of the square profile. A summery of these effects is shown in Fig\,\ref{fig:Figure3} (c).\\
In conclusion diffraction is caused by the pupil of the micro-lenses and interference by the periodicity of the array, and therefore, discrete diffraction orders result according to
\begin{equation} \label{eq:grideq}
    \sin{\left(\varTheta\right)}=m\frac{\lambda}{a}, (m = \pm0,\pm1,...),
\end{equation}
Equation\,(\ref{eq:grideq}) is known as the grid equation in which the grating constant $a$ represents the pitch of the MLA. The difference in optical path length creates an interference pattern with the angular period $\varTheta=\lambda/a$.
The angular spectrum for a standard MLA with a pitch of $\unit[500]{\upmu m}$ and an effective focal length of $\unit[46.5]{mm}$ is shown in Fig.\,\ref{fig:angularspectrum} (parameters see Tab.\,\ref{tab:table1}).
\begin{figure} [t]
  \centering
     \includegraphics[width=0.4 \textwidth]{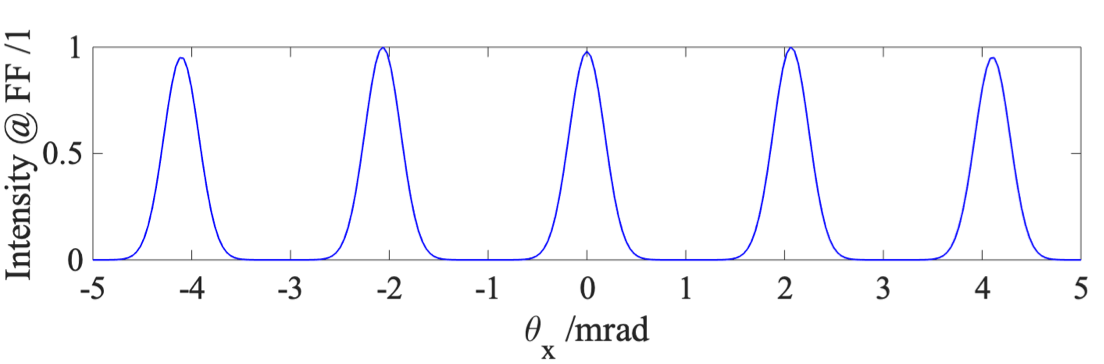}
  \caption{Simulated normalized intensity as function of the diffraction angle for beam splitting based on Fig.\,\ref{fig:Figure3} (c).}
  \label{fig:angularspectrum}
\end{figure}
In addition to the diffraction pattern in the far field, another diffraction pattern is achieved in the focal plane of the first MLA. This diffraction pattern is shown in Fig.\,\ref{fig:beam_MLA_f} (red lines) for the same MLA as used in Fig.\,\ref{fig:angularspectrum}. Due to the lenses, the incident beam is divided into several partial beams (blue lines). According to geometrical optics, the focus result in the center of the micro-lenses. Here the focus represents the main maximum. Moreover, diffraction effects also cause side maxima between the main maxima, which also is visible in Fig.\,\ref{fig:beam_MLA_f}. Obviously, the diffraction pattern in the focal plane of the MLA is dependent on the characteristics of the MLA which includes the effective focal length\,$f\textsubscript{MLA}$ and the numerical aperture $NA=a/(2 \cdot f\textsubscript{MLA})$.\\
As can be seen in Fig.\,\ref{fig:beam_MLA_f}, constructive interference is obtained in the center of each micro-lens. Therefore, the diffraction angle in the focal plane of the MLA\,$\varTheta\textsubscript{MLA}$ results in $2\cdot NA$. Hence for the angle $\varTheta\textsubscript{MLA}$ applies 
\begin{equation} \label{eq:thetamax}
      \varTheta\textsubscript{MLA}=2 NA=\frac{a}{f\textsubscript{MLA}}.
\end{equation} 
Thus, if $\varTheta\textsubscript{max}$ is inserted in \eqref{eq:grideq} for $\varTheta$ and is rearranged to $m$ it follows
\begin{equation}\label{eq:5}
   m = \frac{a^2}{\lambda f\textsubscript{MLA}}, (m=0, 1,...).
\end{equation}
Here, $m$ is the total number of created orders in the far field of the MLA, which is shown in Fig\,\ref{fig:angularspectrum}. It follows that by choosing the focal length\,$f\textsubscript{MLA}$ and the pitch\,$a$, the number of orders\,$m$ can be set. Therefore, in \eqref{eq:5} $m$ represents the number of beams N and the number of channels, respectively, and agrees well with the findings in \cite{streibl1991array, harder2004homogenization, prossotowicz2020dynamic}.
\begin{figure} [t]
  \centering
     \includegraphics[width=0.45 \textwidth]{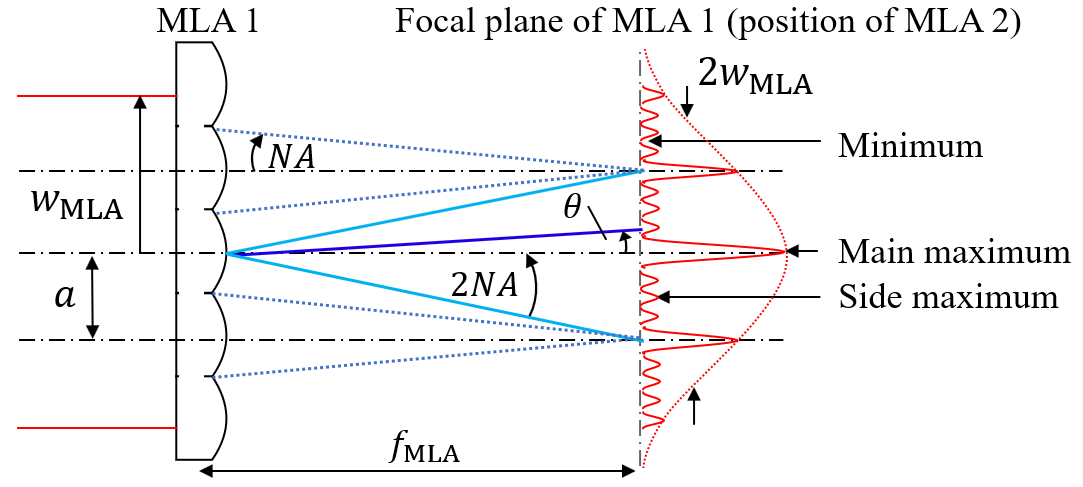}
  \caption{Schematic beam propagation (geometrical optics) for the beam propagation through one MLA until the focal plane.}
  \label{fig:beam_MLA_f}
\end{figure}
\begin{table}
\small
\begin{minipage}[b]{0.9\linewidth}
    \centering
     \caption{\bf Beam splitting and combining parameters.}
    \begin{tabular}{c c c c c c c}
        \hline %
        \(\lambda\textsubscript{0} \)&\( N \)&\( f\textsubscript{MLA}\) & \( a \) & \( f\textsubscript{FL}
        \) & \( FF \) & \( \Delta x\) \\
        \hline %
                $\unit[1.03]{\si\micro m}$&$5\times5$&$\unit[46.5]{mm}$& $\unit[500]{\upmu m}$& $\unit[0.5]{m}$& $\unit[34]{\%}$& $\unit[1]{mm}$\\
       \hline 
    \end{tabular}\label{tab:table1}
    \end{minipage}
\end{table}  
This theoretical consideration, \eqref{eq:thetamax} and \eqref{eq:5}, enable to design an MLA setup acting as beam splitter. With the correct choice of fundamental design parameters an, in particular, equal power distribution can be achieved. Accordingly, the design parameters of MLAs are selected such that an odd or even number of beams $N$ is obtained. A setup where a well defined beam splitting process is possible should be reversible. For the combining process this means that if $N$ channels should be combined the parameters for the MLAs have to fulfill \eqref{eq:5} to reach a high combining efficiency.\\
For beam splitting, the generated beams result in the far field of the first MLA as shown in Fig.\,\ref{fig:Figure3}. Accordingly, for beam combining it is necessary  to arrange the beams in the near field next to each other. This step is identical to the tiled aperture geometry. In contrast to the tiled aperture approach, in the far field a central combining element is placed, which is equal to the filled aperture geometry. Thus, in the plane of the combination element an interference pattern results. The interference condition results from the pitch of the MLAs. Therefore, \eqref{eq:5} must be fulfilled to create a single spot. For this reason, the $FF$ does not limit the combining efficiency as the tiled aperture approach does. This is an advantage compared with the tiled aperture approach, because high combining efficiencies -above \unit[90]{\%}- are possible, independent of the selected $FF$.\\ After the combining element, the combined beam results in the far field. Hence, the MLA combining setup is neither labeled as a tiled aperture nor as filled aperture. For this reason, the term mixed aperture approach is introduced for this new combination method and the advantages of both combination geometries are combined. It is possible to reach high combining efficiencies above $\unit[90]{\%}$ but only one central combining element is used. The mixed aperture is shown in Fig.\,\ref{fig:combination_geometry} (c).\\
The next step for beam combination is to determine the relative phase for each channel to achieve constructive interference. This creates a suitable phase front for the beam combination and ensures a high combination efficiency.
Therefore, the angle spectrum of the beam homogenizer illuminated by the wave function $\psi$ is considered
\begin{align}
    \label{eq:6} \nonumber
   \Psi\left(\varTheta\right)=&\exp{\left(\imath\frac{2\pi}{\lambda} f_{\text{MLA}}\varTheta^2\right)}
   \sum_{m}\exp{\left(\imath \frac{2\pi}{\lambda} f_{\text{MLA}} \frac{\lambda}{a}m \varTheta\right)}\\
   &\tilde{\psi}\left(\varTheta -\frac{\lambda}{a} m\right)
\end{align} 
with $\tilde{\psi}(\varTheta)\sim\exp{\left(-\frac{\varTheta^2}{\varTheta\textsubscript{R}^2}\right)}$ the divergence $\varTheta\textsubscript{R}$ of the incident beams. For details see Ref.\,\cite{tillkorn2018anamorphic}. In \eqref{eq:6} the phase factor accounts for the difference in optical path between the apertures.
The variable $m$ represents the position of the individual beam which should be combined. The split and combined beams are defined as diffraction orders, where the central spot represents the zero order. By this declaration, variable $m=\pm(N-1)/2$ is defined accordingly. Furthermore, the characteristics (position and size) of the individual beams which should be combined play an important role. Therefore, with the use of a FL with the focal length $f\textsubscript{FL}$ (note $f\textsubscript{FL}=f\textsubscript{FL\textsubscript{in}}=f\textsubscript{FL\textsubscript{out}}$) a spacing between the spots in the focal plane follows with $\Delta x = f\textsubscript{FL}\cdot\lambda/a$.\\
Accordingly, the $FF$ can be defined as $FF=2w\textsubscript{in}/\Delta x$ with $w\textsubscript{in}$ the beam waist of the individual beams. For the combined beam the beam waist $w\textsubscript{out}$ depends on the FL which is used to image the far field with the focal length $f\textsubscript{FL}\textsubscript{out}$ and it follows  $w\textsubscript{out}=w\textsubscript{in}\cdot f\textsubscript{FL}\textsubscript{out}/f\textsubscript{FL}$. The beam waist directly at the position of the first MLA $w\textsubscript{MLA}$ is given by $ w\textsubscript{MLA}=2a/(FF \cdot \pi)$.\\
For beam splitting $w\textsubscript{MLA}$ represents the input beam and for beam combination it corresponds to the beam radius at the 2.\,MLA.\\ 
To determine the splitting\,$\eta \textsubscript{split}$ and combining efficiency\,$\eta \textsubscript{comb}$ -not to be confused with the system efficiency- \cite{klenke2016performance}, the intensity distribution of the far field is considered. All undesired orders are considered as efficiency loss, which is referred to boundary fields. This boundary field is cut off and set in relation to the 0.\,Order. This results in the splitting efficiency\,$\eta\textsubscript{split}$ with
\begin{equation}\label{eq:StrahlteilungsEffizienz}
  \eta\textsubscript{split} = \frac{\int\limits_{-D\textsubscript{FF\textsubscript{x}}/2}^{{D\textsubscript{FF\textsubscript{x}}}/2}\int\limits_{-D\textsubscript{FF\textsubscript{y}}/2}^{{D\textsubscript{FF\textsubscript{y}}}/2}I(x,y)dxdy}{\int\limits_{-\infty}^{\infty}\int\limits_{-\infty}^{\infty}I(x,y)dxdy}.
\end{equation}
Here, $I(x,y)$ is the intensity distribution in the far field. The combination efficiency\,$\eta\textsubscript{comb}$ results in
\begin{equation}\label{eq:KombinationsEffizienz}
  \eta\textsubscript{comb} = \frac{\int\limits_{-{\Delta x\textsubscript{comb}}/2}^{{\Delta x\textsubscript{comb}}/2}\int\limits_{-{\Delta y\textsubscript{comb}}/2}^{{\Delta y\textsubscript{comb}}/2}I(x,y)dxdy}{\int\limits_{-\infty}^{\infty}\int\limits_{-\infty}^{\infty}I(x,y)dxdy}.
\end{equation}
\begin{figure*} [t]
  \centering
     \includegraphics[width=0.9 \textwidth]{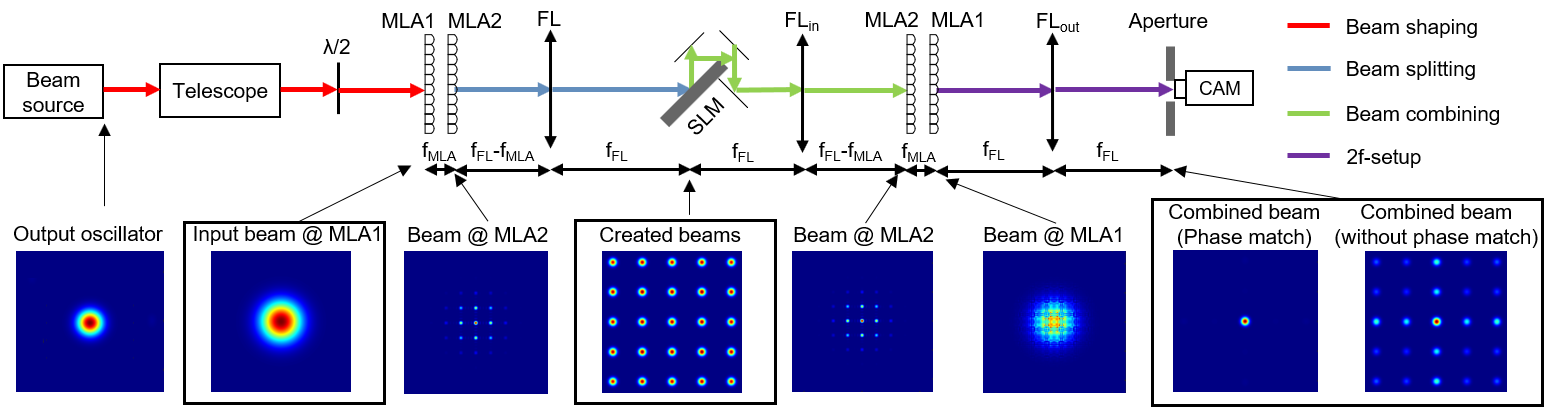}
  \caption{Setup with MLAs as splitting and combining element. Red: beam shaping, blue: beam splitting, green: beam combining, purple: 2f-setup. Additional the simulations for different planes are shown.}
  \label{fig:setup}
\end{figure*}
\begin{figure} [t]
  \centering
     \includegraphics[width=0.46\textwidth]{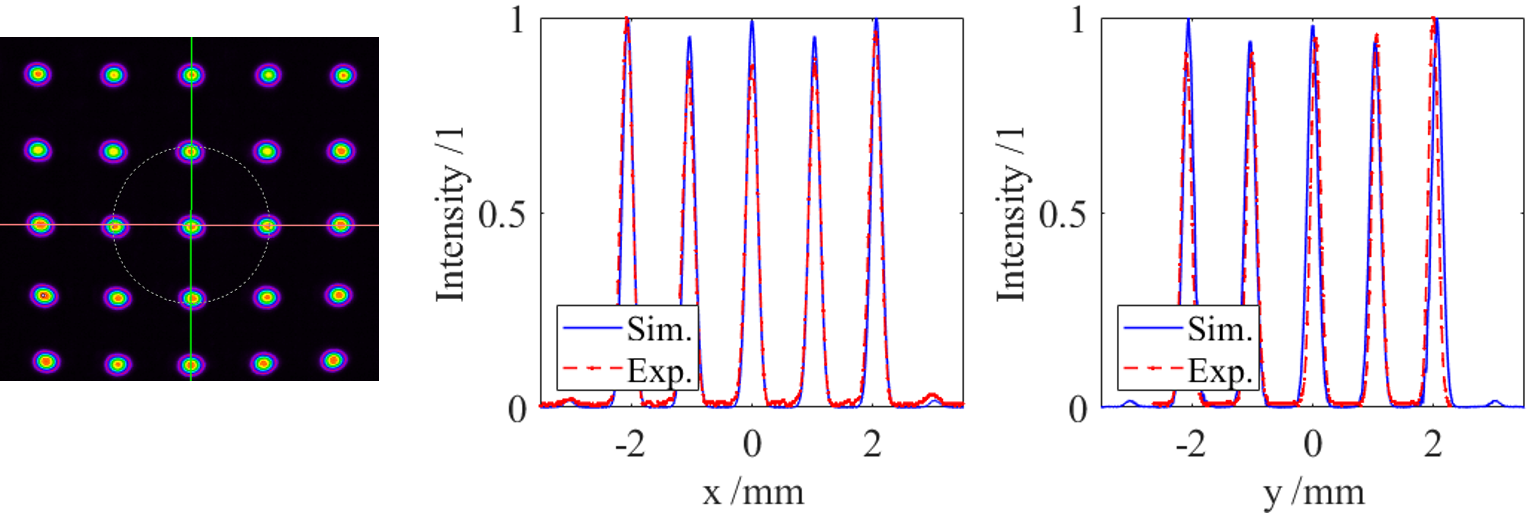}
  \caption{Experimental results for the 2D beam splitting. Image of the generated beam matrix (left). Horizontal (center) and vertical (right) cross-section.}
  \label{fig:Figure8}
\end{figure}
In a proof of principle experiment the beam splitting and combination is done with MLAs (for specifications, see Tab.\,\ref{tab:table1}).
\begin{figure} [t]
  \centering
     \includegraphics[width=0.46\textwidth]{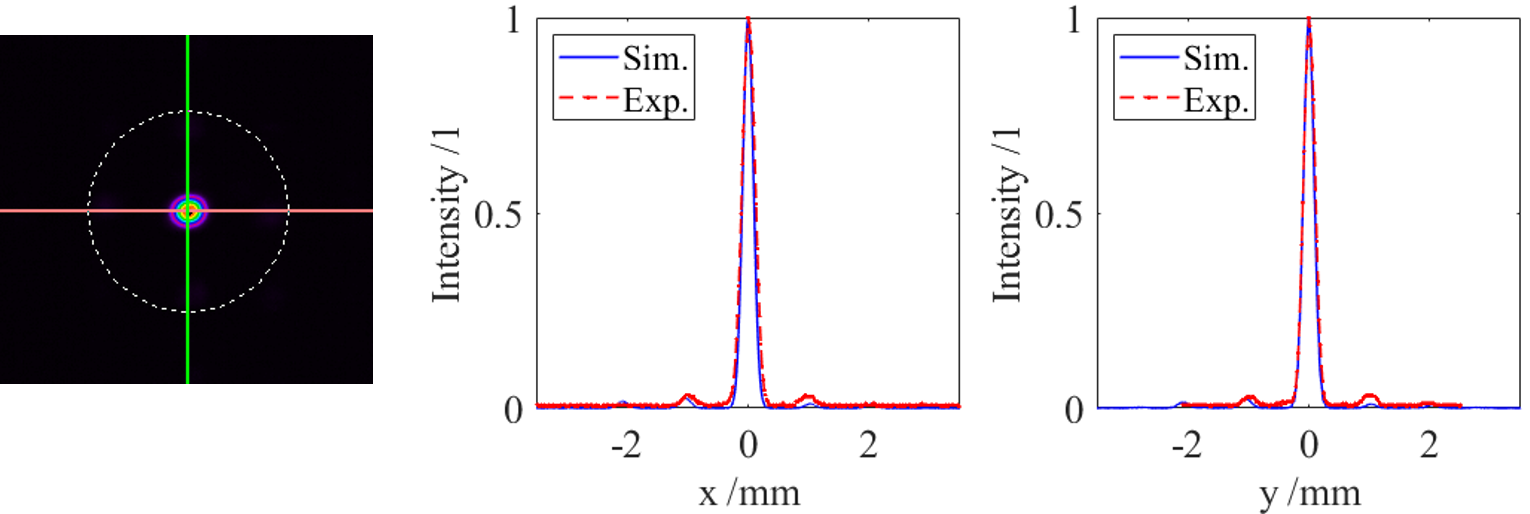}
  \caption{Experimental results for the 2D beam combination. Combined beam with\,/\,without phase match (see \href{https://figshare.com/articles/media/2D_Coherent_beam_combining/13083590}{Visualization 1}). Horizontal (center) and vertical (right) cross-section.}
  \label{fig:Figure10}
\end{figure} 
The experimental setup and simulations are depicted in Fig.\,\ref{fig:setup}. Here, a mode-locked ultra-short pulse laser source at \unit[1030]{nm} with a bandwidth of \unit[10]{nm} is used. Due to the low NA of our setup, a homogeneous intensity distribution results despite the high bandwidth. 
Then a telescope follows that collimates the beam (red line) and a $\lambda/2$-wave plate is used to choose the correct polarization for the SLM. After this the incident Gaussian beam, with a beam waist\,$w\textsubscript{MLA}$ of \unit[0.94]{mm} at MLA1 (approx. 4 micro-lenses are illuminated) is split into small high intensity beamlets, as shown in Fig.\,\ref{fig:beam_MLA_f}. 
In the far field these beamlets interfere and yield a homogeneous distribution of equidistantly $5\times 5$ spaced spots (blue line). 
The phase match, is done with a liquid-crystal-on-silicon based SLM that is positioned in the focal plane of the FL. The SLM represents only one option to control the absolute phase, other phase shifters are also possible. 
Therefore, with the phase factor in \eqref{eq:6} the phase\,$\delta\varphi$ results in
\begin{equation}
    \label{eq:12} 
   \delta\varphi(m\textsubscript{x},m\textsubscript{y})=-\frac{2\pi}{\lambda} f\textsubscript{MLA} \left(\varTheta\textsubscript{x}^2 m\textsubscript{x}^2+\\ 
   \varTheta\textsubscript{y}^2 m\textsubscript{y}^2 \right),  
\end{equation}
valid for Fig.\,\ref{fig:setup}. It should be noted that to calculate the phase a factor of two has to be considered. This is necessary because the phase difference results for the beam splitting and combination process.\\ 
The next step is the beam combination (green line). Here the combination path is the backwards version of the beam splitting part. Consequently, beam profiles at the individual planes, MLA2 and MLA1, agree with the beam profiles in the beam splitting path. At the end, the combination efficiency is proven in the far field. Therefore, a 2f-setup is added (purple line). 
In the plane of the camera we present two cases: once the phases of individual beams at the SLM position are adjusted according to \eqref{eq:12} which leads to an efficient combination (phase match) and in the second case the phases are not adjusted leading to an inefficient combination of the beams (without phase match). \\ 
The result for the beam splitting is presented in Fig.\,\ref{fig:Figure8} and an efficiency\,$\eta\textsubscript{split}$ above $\unit[96]{\%}$ is reached. The aperture effects of the micro-lenses imprint on the side lobes next to the $5\times5$ beam matrix and mainly reduce the splitting efficiency. The barrel distortion in Fig.\,\ref{fig:Figure8} results from the camera setup. 
In Fig.\,\ref{fig:Figure8}, experimental results (red dotted curve) can be directly compared with simulations (blue curve). Therefore, a slight difference in the intensity distribution is visible, which results through the imperfect focal length of the MLA (see \eqref{eq:5}).
\\The beam combination of the generated $25$ beams is depicted in Fig.\,\ref{fig:Figure10} and a combination efficiency of $\unit[90]{\%}$ is achieved. Thus, the efficiency of the experiment is close to the theoretical limit (based on the simulation in Fig.\,\ref{fig:Figure8}) at $\unit[93]{\%}$, which has been determined by simulation, based on \eqref{eq:KombinationsEffizienz}. At this point, it should be noted that no active stabilization is necessary for this low power experiment, because all beams are split from the same source. The situation will be different with independent amplifiers, where an active stabilization is essential to achieve a high combination efficiency.\\
In conclusion, a  novel, compact and simple setup based on MLAs for CBC of $N\times N$ beams is presented. To the best of our knowledge this was the first time that CBC with MLAs was presented. This method is a mixture of the combining geometry tiled and filled aperture. Therefore, the term mixed aperture is introduced. The single channels are placed side by side in the near field and in the far field a central combining element is placed. 
For the combining element a pair of well defined MLAs is used. In the present setup an input beam was split into $5 \times 5$ beams. Subsequently, these $25$ beams were combined to a single beam with a combination efficiency of $\unit[90]{\%}$. Finally, it should be noted, that the presented method offers a simple scaling in terms of the number of combined channels, and, thus for energy and power scaling of a potential high power CBC platform.\\ 
\\
\textsf{\textbf{Disclosures.}} The authors declare no conflicts of interest.

\bibliography{Lit}

\end{document}